\documentstyle[12pt,epsf]{article}

\def\tr{{\hbox{\rm Tr}}}
\def\ex{{\hbox{\rm e}}}

\newcommand{\CS}{{\scriptstyle {\rm CS}}}

\newcommand{\kk}{{{\mathcal K}}}

\newsavebox{\dosii}
\savebox{\dosii}(10,3){
\put(2,5) {\circle{10}}
\put(-3,5){\line(3,0){10}}
\put(2,0){\line(0,5){10} } }
\newcommand{\cdosii}{\mbox{\usebox{\dosii} \hskip 15pt  } }

\begin{document}
\begin{center}

{\large \bf  Knot Invariants and Chern-Simons Theory\footnotemark} 
\vskip5pt

\footnotetext{Invited lecture delivered at the Third European Congress of
Mathematics held at Barcelona, Spain, July 10 -- 14, 2000.}

\vskip1cm

{\large J. M. F. Labastida}

\vspace{0.5cm}

{\em Departamento de F\'\i sica de Part\'\i culas \\
Universidade de Santiago de Compostela \\
E-15706 Santiago de Compostela, Spain\\
{\rm e-mail: labasti@fpaxp1.usc.es}}

\end{center}

\setcounter{page}{1}

\begin{abstract}
A brief review of the development of Chern-Simons gauge theory since its relation
to knot theory was discovered in 1988 is presented. The presentation is done
guided by a dictionary which relates knot theory concepts to quantum field
theory ones. From the basic objects in both contexts the quantities leading to
knot and link invariants are introduced and analyzed. The quantum field
theory approaches that have been developed to compute these quantities are
reviewed. Perturbative approaches lead to Vassiliev or finite type
invariants. Non-perturbative ones lead to polynomial or quantum group
invariants. In addition, a brief discussion on open problems and future
developments is included.  
\end{abstract}


\vskip2cm

In 1988, Edward Witten established the connection between Chern-Simons gauge
theory and the theory of knot and link invariants \cite{csgt}. Since then the
theory has been intensively studied, making important progress as a result of
the application of standard field theory methods. The development of the theory
of knot and link invariants has been also very impressive in the last fifteen
years and at some stages has occurred parallel to  
Chern-Simons gauge theory. There is a natural  correspondence between both
developments.  This has led to the construction of the dictionary introduced in
\cite{baires}, and reproduced in Table I, which will be used as a guide in this
presentation.

Chern-Simons gauge theory was first analyzed from a non-perturbative point of
view. The original paper by Witten presented a series of non-perturbative
methods which led him to establish the equivalence between vacuum
expectation values (vevs) of Wilson loops and
polynomial invariants like the Jones  polynomial \cite{jones} and its
generalizations. Perturbative studies started one year later and soon their
connection to Vassiliev invariants
\cite{vass,birrev} was pointed out. It turned out that
the coefficients of the perturbative series correspond to these
invariants \cite{barnatan,bilin,singular}.

The perturbative series expansion has been studied for different
gauge-fixings. The first analysis in the covariant Landau
gauge \cite{gmm,natan}
was later extended in a general framework
\cite{alla,alts}, reobtaining
the formulation by Bott and Taubes of their configuration space integral
\cite{bt}. Their integral corresponds precisely to the perturbative series
expansion of the vev of a Wilson loop in Chern-Simons gauge theory in the
Landau gauge. Before the work by Bott and Taubes, Kontsevich
presented a different integral \cite{kont} for Vassiliev invariants which
turned out to correspond to the perturbative series  expansion of the vev of a
Wilson loop in the light-cone gauge \cite{cata,lcone,kaucone}.

Additional studies of the perturbative series expansion have been performed in
the  temporal gauge \cite{cata,temporal}.
This gauge has the important feature that the integrals which are
present in the expressions for the coefficients of the perturbative
series expansion can be carried out, leading to combinatorial
expressions \cite{temporal}. This has been shown to be the case up to order
four and it seems likely that the approach can be generalized. In this
analysis a crucial role is played by the factorization theorem for
Chern-Simons gauge theory proved in \cite{factor}. The resulting expressions
are better presented when written in terms of  Gauss diagrams for knots
\cite{faroknots}. Recent results demonstrate the  existence of a combinatorial
formula of this type \cite{virgpo}. Chern-Simons gauge theory has provided
combinatorial expressions for all the Vassiliev invariants up to order four
\cite{temporal}. Further work is needed to obtain a general combinatorial
expression.

The invariants obtained in the perturbative framework for each gauge-fixing
are the same. This is guaranteed by the fact that the theory is gauge
invariant and Wilson loops are gauge invariant operators. In fact, this
property is the responsible, from a field theory point of view, of the
connection between Vassiliev invariants and polynomial invariants, as they
appear in the non-perturbative approach, and of the existence of different
representations of Vassiliev invariants. The correspondence between the
Chern-Simons gauge theory description and the knot theory one is listed in the
following table.

\begin{center}
{\bf Table I}
\end{center}
\vskip-15pt
\begin{table}[hp]
\begin{center}
\begin{tabular}{||c||c||}
\hline {\bf Knot Theory} & {\bf Chern-Simons Gauge Theory} \\
\hline\hline Knots and links & Wilson loops \\
\hline Knot and link polynomial invariants &Vevs of products of Wilson
loops \\
\hline Singular knots & Operators for singular knots \\
\hline Invariants for singular knots & Vevs of the new operators  \\
\hline Finite type or Vassiliev invariants & Coeffs. of the
perturbative series \\
\hline Chord diagram & First coeff. of the perturbative series \\
\hline \{1T,4T\} and \{1T,AS,IHX,STU\} & Lie-algebra structure of group
factors \\
\hline Configuration space integral & Landau gauge \\
\hline Kontsevich integral & Light-cone gauge \\
\hline ?? & Temporal gauge \\
\hline
\end{tabular}
\end{center}
\end{table}

\vskip-9pt
\noindent Before entering into the review of the developments of the last
years guided by this table it is worth to remark two facts. First, the entry in
the knot-theory column corresponding to the temporal gauge has not been filled
in yet. Further work is needed to complete it. Second, recent developments
based on the application of Maldacena's conjecture has led to introduce a new
context in which knot invariants are organized differently \cite{oova}. It is
possible that some important new boxes are still missing in the table.


Chern-Simons gauge theory on a
smooth three-manifold $M$ is defined by the action,
\begin{equation}
S_{\CS} (A) ={k\over 4\pi} \int_M \tr (A\wedge d A + \frac{2}{3} A\wedge
A\wedge A).
\label{valery}
\end{equation}
where $k$ is an integer constant. The exponential of this action is gauge
invariant.  Wilson loops are some of the relevant operators of the
theory. They are defined as:
\begin{equation}
W_\gamma^R (A) =\tr_R \big( {\hbox{\rm Hol}}_\gamma (A) \big) =
\tr_R {\hbox{\rm P}} \exp \int_\gamma A,
\label{rosalia}
\end{equation}
where $R$ denotes a representation of the gauge group $G$ and $\gamma$ a
1-cycle. Products of these operators are the natural candidates to obtain
topological invariants after computing their vev
respect to the action (\ref{valery}).

The Wilson loop (\ref{rosalia}) and the vevs of its products are the first two
items of the right column of Table I. The first two in the left column are the
basic ingredients of knot theory.  Knot theory studies inequivalent embeddings
$\gamma:$
$S^1
\rightarrow M$. Each of these is a knot. Knots are classified constructing  knot
invariants or quantities which can be computed taking a
representative of a class and are invariant within the class. 
The problem of the classification of knots in $M=S^3$ can be reformulated in a
two-dimensional framework using regular knot projections. The previous
equivalence problem then translates into the equivalence of regular knot
projections under Reidemeister moves.

The study of knot and link invariants experimented important progress in
the eighties after the discovery of the Jones polynomial
\cite{jones} and its generalizations like the HOMFLY
\cite{homfly} and the Kauffman \cite{kauffman} polynomial invariants. 
Witten showed in 1988 that the vevs of products of Wilson
loops correspond to the Jones polynomial when one considers
$SU(2)$ as gauge group and all the Wilson loops entering in the vev are taken
in the fundamental representation $F$. For example, if one considers a knot
$K$, with Jones polynomial $V_K(t)$,  he showed that,
$V_K(t)= \langle W_{K}^{F} \rangle$, provided that one performs the
identification $ t= \exp({2\pi i / k+h})$
where $h=2$ is the dual Coxeter number of the gauge group $SU(2)$. He
also showed that if instead of $SU(2)$ one considers $SU(N)$ and the Wilson
loop carries the fundamental representation, the resulting invariant is the
HOMFLY polynomial. If, instead, one considers $SO(N)$ as
gauge group and  Wilson loops carrying the fundamental representation one is
led to the Kauffman polynomial. The case of $SU(2)$ as gauge
group and Wilson loops carrying a representation of spin $j/2$ leads to the
Akutsu-Wadati \cite{aku} polynomials. The framework generated by Chern-Simons
gauge theory leads to an enormous variety of knot and link invariants.  They
can be also obtained from a quantum group approach \cite{qga}, and from more
general formalisms \cite{cate}.

In our discussion of the next five items in Table I we will deal first with
the left column. In 1990, V. A.  Vassiliev \cite{vass} introduced  new
knot invariants based on singular knots which were reformulated later
by Birman and Lin \cite{bilin} from
an axiomatic point of view. A   singular knot
with $j$ double points consists of the image of a map from $S^1$ into $S^3$
with $j$ simple self-intersections.  The key ingredient in the construction by
Birman and Lin is the observation that any knot invariant extends to generic
singular knots by the Vassiliev resolution:
\begin{equation}
\nu({K^{j+1}}) =\nu({K^j_+}) - \nu({K^j_-}),
\label{vassreso}
\end{equation}
where $K^{j+1}$ is a singular knot with $j+1$ double points which differs
from the knots $K^j_+$ and  $K^j_-$ only in the region where the double point
is resolved by an overcrossing $(+)$ and an undercrossing $(-)$. Using this
extension Birman and Lin \cite{bilin} characterized the invariants of finite
type or Vassiliev invariants introducing the following definition: a Vassiliev
or finite type invariant of order $m$ is a knot invariant which is zero on the
unknot and that, after extending it to singular knots, it is zero on singular
knots $K^j$ with $j>m$ double points.

Besides introducing an axiomatic approach to Vassiliev invariants, Birman and
Lin proved an important theorem in 1993 \cite{bilin}. Any
polynomial invariant $P_K(t)$ for a knot $K$ can be expanded as:
\begin{equation}
Q_K(x) = P_K(\ex^x) = \sum_{m=0}^\infty \nu_m (K) x^m.
\label{elteo}
\end{equation}
Birman and Lin proved that if one extends the quantities $\nu_m(K)$ to
Vassiliev invariants for singular knots using Vassiliev resolution
(\ref{vassreso}), then $\nu_m(K)$ are Vassiliev invariants of order $m$. An
immediate consequence of this theorem is that the coefficients of the
perturbative expansion associated to the vev of a Wilson loop in Chern-Simons
gauge theory are Vassiliev invariants. This property of the coefficients of
the perturbative series expansion has been proved using standard quantum
field theory methods \cite{singular}.

From a singular knot with $m$ double points one can construct a
particular object which determines Vassiliev invariants of order
$m$: its chord diagram \cite{barnatan}. Given a singular knot $K^m$, its chord
diagram, $CD(K^m)$, is built in the following way. Take a base point and draw
the preimages of the map associated to a given representative of $K^m$ on a
circle. Then join by straight lines the pairs of preimages which correspond to
each singular point. If $\nu(K^m)$ is a Vassiliev invariant of
order $m$ then it is completely determined by $CD(K^m)$.

Chord diagrams play an important role in the theory of Vassiliev invariants
\cite{barnatan}. Since Vassiliev invariants of order $m$ for singular knots with
$m$ double points are codified by chord diagrams one could ask if there are as
many independent invariants of this kind as chord diagrams. The answer to this
question is no. Chord diagrams are associated to knot diagrams
and these diagrams must be considered modulo the equivalence relation
dictated by the Reidemeister moves. These relations indeed
impose some relations among chord diagrams, the so-called 1T and 4T relations
\cite{barnatan}. The general expression for the dimensions of the spaces  of chord
 diagrams is an open problem which has challenged many
people. These dimensions correspond in fact to
the dimensions of the spaces of primitive Vassiliev invariants.

The vector space of chord diagrams can be characterized in an
equivalent way using trivalent diagrams an introducing a series of new
relations. This characterization is very important because it corresponds to
the one that naturally arises from the point of view of Chern-Simons gauge
theory. It consists of the expansion of the set of chord diagrams to a new
set in which trivalent vertices are allowed. This means that now the lines in
the interior of the circle can join a point on the circle to a point on one of
the internal lines. 
Bar-Natan showed
\cite{barnatan} that the space of chord diagrams modulo 1T and 4T relations is
equivalent to the new one after modding out by the so-called  1T, AS, IHX and
STU relations.

The relations AS, STU and IHX are reminiscent of a Lie-algebra structure. If
one assigns totally antisymmetric structure constants $f_{abc}$ to the
internal trivalent vertices, and group generators $T_a$ to the vertices on the
circle, the STU relation is just the defining Lie-algebra relation,
while the IHX relation corresponds to the Jacobi identity.
The group factors associated to the perturbative
series expansion of the vev of a Wilson loop in Chern-Simons gauge theory
correspond precisely to these spaces. 

We will now turn our attention to the column on the right column in
Table I. Singular knots play a central role in the theory of Vassiliev
invariants. As shown in \cite{singular}, they have an operator counterpart
in Chern-Simons gauge theory. It has a rather simple form. 
Let us consider a singular knot
 $K^n$ with $n$ double points, and let us assign to
each double point $i$ a triple $\tau_i=\{s_i,t_i,T^{a_i}\}$ where $s_i$ and
$t_i$, $s_i<t_i$, are the values of the $K^n$-parameter at the double
point, and $T^{a_i}$ is a group generator. The gauge-invariant operator
associated to the singular knot $K^n$ is:
\begin{eqnarray} && {\hskip-2cm} ({4\pi i \over k})^n \tr \Big[T^{\phi(w_1)}
U(w_1,w_2) T^{\phi(w_2)} U(w_2,w_3) T^{\phi(w_3)}\cdots \nonumber\\ &&
{\hskip3cm}
\cdots
U(w_{2n-1},w_{2n}) T^{\phi(w_{2n})} U(w_{2n},w_{1}) \Big],
\label{operador}
\end{eqnarray} 
where $\{w_i;i=1,\dots,2n\}$, $w_i<w_{i+1}$, is the set that
results from ordering the values $s_i$ and $t_i$, for $i=1,\dots,n$, and
$\phi$ is a map that assigns to each $w_i$ the group generator in the
triple to which it belongs.

Some immediate implications of the singular operators (\ref{operador}) are the
following. First, they lead to a proof of the
theorem by Birman and Lin discussed above; second, they allow to make direct
contact with chord diagrams since these diagrams correspond to these operators
at lowest order. This has been shown in \cite{singular}. The quantities which
result after the assignment of Lie-algebra data to chord diagrams are called
weight systems
\cite{barnatan}. For each system one chooses a group and a representation. They
correspond to the group theory factors in the context of Chern-Simons
perturbation theory. 

We will now describe the last three items of Table I starting with the right
column. Perturbative studies of Chern-Simons gauge theory started with the
works by  Guadagnini, Martellini and Mintchev
\cite{gmm} and by Bar-Natan \cite{natan}. These studies were made in the
covariant Landau gauge. Subsequent works
\cite{alla,alts} in this gauge led to a framework linked to the theory of
Vassiliev invariants, which constituted the configuration space integral
approach \cite{bt}. The elements of the resulting Feynman rules in this
gauge are:
\begin{equation}
{i\over 4\pi} \delta_{ab}\epsilon^{\mu\nu\rho}{(x-y)_\rho
\over |x-y|^3}, \,\,\,\,\,\,\,\,
-i g f_{abc} \epsilon_{\mu\nu\rho} \int d^3x , \,\,\,\,\,\,\,\,
g (T^a)_i^j.
\label{nuevetres}
\end{equation}
which correspond to
internal line (gauge propagator), internal vertex, and vertex on the Wilson
loop, respectively. In these equations $g^2=4\pi/k+h$. To these rules one must
include the ones related to the ghost fields present in the Landau gauge. One
of the consequences of their presence is that higher-loop corrections to two-
and three-point functions can be ignored, at least in some of the standard
regularization schemes.

In analyzing the perturbative series one must deal with an important subtlety.
If one computes the first order contribution to the perturbative series
expansion of the vev of a Wilson loop one finds that the resulting quantity is
not a topological invariant. In the gauge fixing of the theory we have
introduced a metric dependence that could lead to quantities which are not
topological. This first order contribution is just a manifestation of it.
Fortunately, only in this term, and in its propagation in higher order
contributions, topological invariance is lost. The rest of the perturbative
series expansion is truly topological. Thus, although vevs are not topological
invariant quantities, they fail to be so in a controllable way. The
non-topological terms factorize and multiply a term which is topological. The
factor turns out to have a framing dependence equivalent to the one obtained in
non-perturbative approaches.

 The Feynman rules allow to split the
contributions to each order in two factors: a geometrical factor which
includes all the space dependence, and  a group factor which includes all the
group theoretical dependence. The general form is:
\begin{equation}
\langle W^R_K \rangle = \hbox{\rm dim}\, R \sum_{i=0}^\infty
\sum_{j=1}^{d_i}\alpha_{ij}(K) r_{ij}(R) x^i,
\,\,\,\,\,\,\,\,\,\,\,\,\,\,
x={2\pi i\over k+h}=ig^2/2,
\label{expansion}
\end{equation}
where $R$ is the dimension of the representation $R$,  
$\alpha_{0,1}=r_{0,1}=1$, and $d_0=1$.  The factors $\alpha_{ij}(K)$ and
$r_{ij}(R)$ appearing at each order $i$ incorporate all the dependence
dictated from the Feynman rules apart from the dependence on  the
coupling constant, which is contained in $x$. 
Of these two factors, in the $r_{ij}(R)$ all
the group-theoretical dependence is collected. The rest is contained in  the
$\alpha_{ij}(K)$ or geometrical factors. They have the form of integrals over
the Wilson loop corresponding to the knot $K$ of products of propagators, as
dictated by the Feynman rules. The first index in $\alpha_{ij}(K)$ denotes the
order in the expansion and the second index labels the different geometrical
factors which can contribute at the given order. Similarly, $r_{ij}(R)$ stands
for the independent group structures which appear at order
$i$, which are also dictated by the Feynman rules. The object
$d_i$ in (\ref{expansion}) is the dimension of the space of
invariants at a given order. 

Among the basis of group factors which can be chosen  there is a special class 
 called canonical basis which turns out to be very
useful. Basically, it consists of connected diagrams. If we denote by
$r_{ij}^c(R)$ the group factors associated to this basis, and
$\alpha_{ij}^c(K)$ the corresponding geometrical factors, the perturbative series
expansion (\ref{expansion}) can be written as \cite{factor}:
\begin{equation}
\langle W^R_K \rangle = \hbox{\rm dim}\, R \,\exp\left\{ \sum_{i=1}^\infty
\sum_{j=1}^{\hat d_i} \alpha_{ij}^c(K)\,r_{ij}^c(R)\,x^i\right\},
\label{expansiondos}
\end{equation}
where $\hat{d}_i$ stands for the number of connected elements in the
canonical basis at order $i$.  The result
(\ref{expansiondos}) is known as the factorization theorem, and  it holds
for arbitrary gauges.  The geometrical factors 
$\alpha_{ij}^c(K)$ are a selected set of Vassiliev invariants. They are
called primitive Vassiliev invariants.  They have been computed for general
classes of knots as torus knots \cite{torusknots,simon} up to order six.

The contribution at first order in (\ref{expansiondos}) is precisely the
framing factor.  The rest of the terms in
the exponent of (\ref{expansiondos}) are knot invariants. The series
contained in that exponent was analyzed by Bott and Taubes \cite{bt} in their
work on the configuration space for Vassiliev invariants (listed on the left
column in Table I). They showed that the integral expression entering the
geometrical factors
$\alpha_{ij}^c(K)$ are convergent \cite{bt,dylan}. Further work on the subject
has led to a proof of their invariance \cite{alts,yang}.

The explicit expression for the integrals entering in the second order
contribution was first presented in  \cite{gmm}. It was later analyzed in
detail by Bar-Natan
\cite{natan}.
This invariant turns out to be the total twist of the knot and coincides mod 2
with the Arf invariant. The integral expression for the order three
invariant, 
$\alpha_{31}^c(K)$ was first presented in  \cite{alla}. Properties of the 
primitive Vassiliev invariants $\alpha_{21}^c(K)$ and
$\alpha_{31}^c(K)$ have been studied in
\cite{alemanes}. In these works the integral
expressions for $\alpha_{21}^c(K)$ and $\alpha_{31}^c(K)$ were studied in the
flat-knot limit and combinatorial expressions were obtained.

The perturbative analysis of Chern-Simons gauge theory in the
 light-cone gauge leads to the Kontsevich integral, which
constitutes a particular representation of Vassiliev invariants.
Non-covariant gauges are characterized by a unit constant
vector $n$ and have the form $n^{\mu} A_{\mu} = 0$.
In the case of the light-cone gauge the unit vector $n$ satisfies the
condition $n^2=0$. In this gauge there is only one Feynman rule to be taken into
account to compute the vevs of operators: the one associated to the propagator.
The group factors that remain in this case correspond just to chord diagrams.
The fact that in this gauge  no group factors with trivalent vertices have to be
taken into account is a quantum field theory ratification of Bar-Natan theorem
among the equivalence of the two representations of the space of diagrams.
Non-covariant gauges share the problem of the presence of unphysical poles in
their propagators \cite{leibrew}. Several
prescriptions have been proposed to avoid these unphysical poles. Usually, a
prescription is chosen so that some particular properties of the correlation
functions are fulfilled. In the light-cone gauge there is a natural
prescription which is motivated by the simple form that the elements of the
perturbative expansion take after performing a Wick rotation. This prescription
leads \cite{lcone} to the Kontsevich integral.

The studies in the Landau and in the light-cone gauge provide integral
expression for Vassiliev invariants. It is difficult to obtain information on
these invariants from these expressions. Combinatorial formulas are
much preferred.  It is known that a general
combinatorial formula for Vassiliev invariants exists
\cite{virgpo,arrgpo}.   The search for an explicit construction of the 
combinatorial formula has led to the study of Chern-Simons gauge theory in the
temporal gauge \cite{temporal}. This turns out to be the more
suitable gauge to carry out all the intermediate integrals and obtain
combinatorial expressions. This approach has provided a combinatorial
expression for the two primitive Vassiliev invariants at order four. The
temporal gauge has been also treated in \cite{cata,vande}. Previous studies of
the configuration space integrals in the limit of flat knots \cite{alemanes}
have also led to combinatorial expressions for Vassiliev invariants of order
two and three.

The starting point of the analysis in the temporal gauge is the same as in
the light-cone gauge. The gauge-fixing condition is the same but
now $n$ is a unit vector of the form $n^\mu=(1,0,0)$.  As before, the
propagator presents unphysical poles, and a prescription to regulate
it is needed. In this case a prescription-independent analysis is done splitting
the propagators in two terms. It leads to the concept of kernel, as introduced in
\cite{temporal}. The kernels are quantities which depend on the knot projection
chosen and therefore are not knot invariants. However, at a given order $i$ a
kernel differs from an invariant of type $i$ by terms that vanish in signed sums
of order $i$. The kernel contains the part of a Vassiliev invariant which is
the last in becoming zero when performing signed sums, in other
words, a kernel vanishes in signed sums of order $i+1$ but does
not in signed sums of order $i$. 
Kernels plus the structure of the perturbative series expansion
seem to contain enough information to reconstruct the full
Vassiliev invariants \cite{temporal}. The general expression for the kernels can
be written in a universal form much in the spirit of the universal form which
shares some resemblence with Kontsevich integral. 
 
Using the kernels and taking into account general properties of the perturbative
series expansion one can reconstruct the complete perturbative coefficients
obtaining combinatorial formulas. Vassiliev invariants up to
order four were expressed in terms of these quantities and the crossing
signatures in ref. \cite{temporal}.  Here, we collect only the formula for
the primitive Vassiliev invariant at second order. It has the following
expression:
\begin{equation}
\alpha_{21}(K) = \alpha_{21}(U) +  
\langle \cdosii \hbox{\hskip-0.4cm}, \bar
G(\kk) \rangle,
\label{primtwoc}
\end{equation}
where  $\alpha_{21}(U)$ stands for the value of $\alpha_{21}$
for the unknot and $\bar G(\kk) = G(\kk) - G(\alpha(\kk))$, where 
$\alpha({\kk})$ denotes the ascending diagram of the knot projection $\kk$.
$G(\kk)$ is the Gauss diagram corresponding to $\kk$.
The inner product used in (\ref{primtwoc}) consists of the sum over all the
embeddings of the diagram $\cdosii \hbox{\hskip-0.4cm}$ into $G({\kk})$, each
one weighted by a factor,
$
\varepsilon_1 \varepsilon_2,
$
where $\varepsilon_1$ and  $\varepsilon_2$ are the
signatures of the chords of $G({\kk})$ involved in the embedding.

The analysis presented in \cite{temporal} up to order four should
be generalized to arbitrary order, trying to obtain a general expression similar
to the one existing in the light-cone corresponding to the Kontsevich integral.
The resulting formula would allow to fill the last box on the left column of
Table I. Though this study seems promising, the problems inherent to the
proper treatment of gauge theories in non-covariant gauges constitute an
important barrier. Much work has to be done to understand the subtle issues
related to the use of non-covariant gauges. The kernels plus the properties of
the perturbative series expansion are probably enough to compute the explicit
form of a given invariant but certainly it does not provide a systematic way of
deriving the general universal formula.

The relation between knot theory and Chern-Simons gauge theory does not end here.
Most likely  additional boxes to Table I are waiting to be discovered. Quantum
field theory is a very rich framework which is enlarged when regarded from the
point of view of string theory. Recent work \cite{oova} indicates that new
important connections can be established that could lead to entirely new
approaches to the theory of knot invariants.

\end{document}